\documentclass[twocolumn,showpacs,aps]{revtex4}
\usepackage{graphicx}
\usepackage{xcolor}
\usepackage{subfigure}
\def\scr{\rm\scriptscriptstyle }

\begin{document}

\title[]{The nuclear and Coulomb breakup of the weakly bound $^{6}$Li with
targets in the range from A = 59 to 208} 
\author{D.R. Otomar, P. R. S. Gomes, J. Lubian}
\affiliation{Instituto de F\'{\i}sica, Universidade Federal Fluminense, Av. Litoranea
s/n, Gragoat\'{a}, Niter\'{o}i, R.J., 24210-340, Brazil}
\author{L. F. Canto}
\affiliation{Instituto de F\'{\i}sica, Universidade Federal do Rio de Janeiro, CP 68528,
Rio de Janeiro, Brazil}
\author{M. S. Hussein}
\affiliation{Instituto de Estudos Avan\c{c}ados, Universidade de S\~{a}o Paulo C. P.
72012, 05508-970 S\~{a}o Paulo-SP, Brazil, and Instituto de F\'{\i}sica,
Universidade de S\~{a}o Paulo, C. P. 66318, 05314-970 S\~{a}o Paulo,-SP,
Brazil}
\keywords{breakup, elastic scattering}
\pacs{24.10Eq, 25.70.Bc, 25.60Gc }

\begin{abstract}
We have performed CDCC calculations for the $^{6}$Li + $^{59}$Co, $^{144}$Sm
and $^{208}$Pb systems, to investigate the dependence of the relative
importance of nuclear and Coulomb breakup on the target charge (mass) at
near barrier energies. The calculations were in good agreement with the
experimental elastic scattering angular distributions for these systems and
then, their predictions to the nuclear, Coulomb and total breakup were
investigated. Although the relative importance of the nuclear breakup is, as
expected, larger for lighter targets, this effect is not very pronounced. We
also investigate a scaling of the nuclear breakup with the target mass and
we compare the predictions for the integrated total breakup cross sections
with experimental fusion cross sections at similar energies.
\end{abstract}

\maketitle

\section{\protect\bigskip Introduction}

The breakup of weakly bound nuclei has been a subject of great interest
in the last years~\cite{CGD06}, both theoretically and experimentally. 
Among the main investigations within this field, there are studies of the 
breakup cross section and the influence of the breakup process on other 
reaction channels. The conclusions from those investigations may
change with the characteristics of the weakly bound nuclei (from now on we
will assume that they are the projectiles), the targets and energy regime involved. For
example, halo nuclei which have extremely low breakup energy threshold may
behave differently from stable weakly bound nuclei like $^{6}$Li, $^{7}$Li
and $^{9}$Be. Usually, it is assumed that Coulomb breakup predominates over 
nuclear breakup when heavy targets are involved. The situation may be different
in the case of light targets. Thus, the nature of the breakup process should depend
strongly on the target mass. 

Regarding the influence of the breakup process on the fusion
cross sections, at the present there is a general qualitative understanding
that breakup enhances fusion at sub-barrier energies, whereas it produces 
some suppression above the barrier~\cite{CGL09,Gomes09}. 
Concerning the energy dependence of the optical potential
in the scattering of weakly bound systems, several works show a behavior
different from the one found for tightly bound systems. This behaviour is usually
called the {\it breakup threshold anomaly} (BTA) \cite{Hussein}. Recently it
has been shown experimentally that transfer processes followed by breakup
may predominate over direct breakup of stable weakly bound nuclei at
sub-barrier energies~\cite{Luong,Dasgupta10,Rafiei, Shiravasta}. 

In the present work we investigate the breakup process evaluating separate 
contributions from the Coulomb and from the nuclear fields,
as well as the Coulomb-nuclear interference. We perform calculations for collisions of
$^{6}$Li projectiles with $^{59}$Co, $^{144}$Sm and $^{208}$Pb, at near-barrier energies. 
Since the target's atomic numbers are 27, 62 and 82, one may consider $^{6}$Li + $^{59}$Co, 
$^{6}$Li + $^{144}$Sm and $^{6}$Li + $^{208}$Pb as a light system, a medium system 
and a heavy system, respectively. The choice of these systems was based on the availability
of elastic scattering data in the literature. In this way, we could check the reliability of
our calculations through comparisons with the scattering data.

The present paper is organised as follows. In section II we present the theoretical method 
used in the calculations. In sections III to V we show the results and discuss them. Finally, 
in section VI we present the conclusions of our work. \bigskip 

\section{CDCC calculations}

It is widely accepted that the most suitable approach to deal with the
breakup process, which feeds states in the continuum, is the so called
continuum discretized coupled channel (CDCC) method \cite{KYI86,AIK87}.
In this type of calculations, the continuum wave functions are grouped in
bins or wave packets that can be treated similarly to the usual bound
inelastic states, since they are described by square-integrable wave
functions. In the present work we use the same assumptions and methodology
adopted in the CDCC calculations of Ref.~\cite{Otomar10}, where the elastic
scattering of the $^{6}$Li + $^{144}$Sm system was investigated. We present
below a summary of the main points of the CDCC method. Further details of the
calculations can be found in Refs.~ \cite{Otomar10,KYI86,AIK87}.

Collisions involving $^{6}$Li projectiles are influenced by the continuum states, 
representing mainly the breakup of $^{6}$Li into a deuteron and an alpha particle. Owing to the 
low threshold of this breakup reaction (S$_{\alpha}=1.47$ MeV), the breakup channel
is strongly coupled with the elastic channel. Thus, it is necessary to include the
continuum in the coupled channel calculation and this can be done with the CDCC method.
For this purpose, we use the cluster model in which $^{6}$Li is described as a bound 
state of the $d+\alpha$ system and the breakup channel is represented by the continuum 
states of this system. This model has been successfully used in previous CDCC calculations
in collisions of $^{6}$Li projectiles~\cite{SYK82,KRu96}. The numerical calculations were
performed using the computer code FRESCO~\cite{Tho88}. In the cluster model, the 
projectile-nucleus interaction is written as

\begin{equation}
V(\mathbf{R},\mathbf{r},\xi) = V_{\alpha-{\scr T}}(\mathbf{R},\mathbf{r},\xi)
+ V_{d-{\scr T}}(\mathbf{R},\mathbf{r},\xi)
\label{potentials}
\end{equation}
where {\bf R} is the vector joining the projectile's and target's centers,  
 {\bf r} is the relative vector between the two clusters ($d$ and $\alpha$),
 and $\zeta $ stands for any other intrinsic coordinate describing the projectile-target
 system. 
 
 In our calculations, the continuum states of $^{6}$Li are discretised as in 
 Refs.~\cite{DTB03,OLG09,Otomar10}. Thus, we do not repeat the details here.  
 The interaction between the $d$ and the $\alpha$ clusters within $^{6}$Li
 is given by a  Woods-Saxon potential, with the same parameters as in
 Refs.~\cite{DTB03,OLG09,Otomar10}.

The real parts of the $V_{\alpha-{\scr T}}(\mathbf{R},\mathbf{r},\xi)$
and $V_{d-{\scr T}}(\mathbf{R},\mathbf{r},\xi)$ interactions were given
by the double-folding S\~ao Paulo potential~\cite{Chamon}. We have 
assumed that the mass densities of the $d$ and $\alpha$ clusters,
required for the double-folding calculation,
can be approximated by the charge densities multiplied by two, whereas the
mass densities of the $^{59}$Co, $^{144}$Sm and $^{208}$Pb targets
were taken from the systematic study of Ref.~\cite{Chamon}. The imaginary
parts of $V_{\alpha-{\scr T}}(\mathbf{R},\mathbf{r},\xi)$ and 
$V_{d-{\scr T}}(\mathbf{R},\mathbf{r},\xi)$ were chosen as to
represent short-range fusion absorption. They were given by 
Woods-Saxon functions with depth $W_0 = -50$ MeV, radial parameter
$r_{0i} = 1.06$ fm and diffusivity $a_i = 0.2$ fm. These imaginary potentials
correspond to taking ingoing wave boundary conditions.

The present CDCC calculations include also inelastic channels, corresponding to
collective excitations of the targets. These channels were selected according to the 
specific nuclear structure properties of the target. For $^{144}$Sm, the
excitations included were the one-phonon quadrupole (2$^{+}$, E$^{\ast }$ =
1.660 MeV) and octupole (3$^{+}$, E$^{\ast }$ = 1.8102 MeV) first order
vibrations. The values of the deformation parameters were obtained from Ref.~\cite{Ram01} 
and \cite{Kib02} for the quadrupole and octupole deformations,
respectively. For the $^{208}$Pb target, we consider collective 3$^-$ (E$^{\ast }$ = 2.6145 MeV) 
and 5$^-$ (E$^{\ast }$ = 3.1977 MeV) states. The deformation parameters were taken 
from Ref. \cite{pb208}.  As for the $^{59}$Co target the quintuplet of identified states 
associated with the $2^+$ collective excitations were approximated by a single level  with the 
energy equal to the centroid of the multiplet and the deformation length corresponding to that of
the combined states \cite{co59}. 

\section{Results of the CDCC calculations}

\begin{figure}[ht]
\centering
 \begin{minipage}[ht] {1\linewidth}
 \centering
 \subfigure{\includegraphics[width=0.8\linewidth]{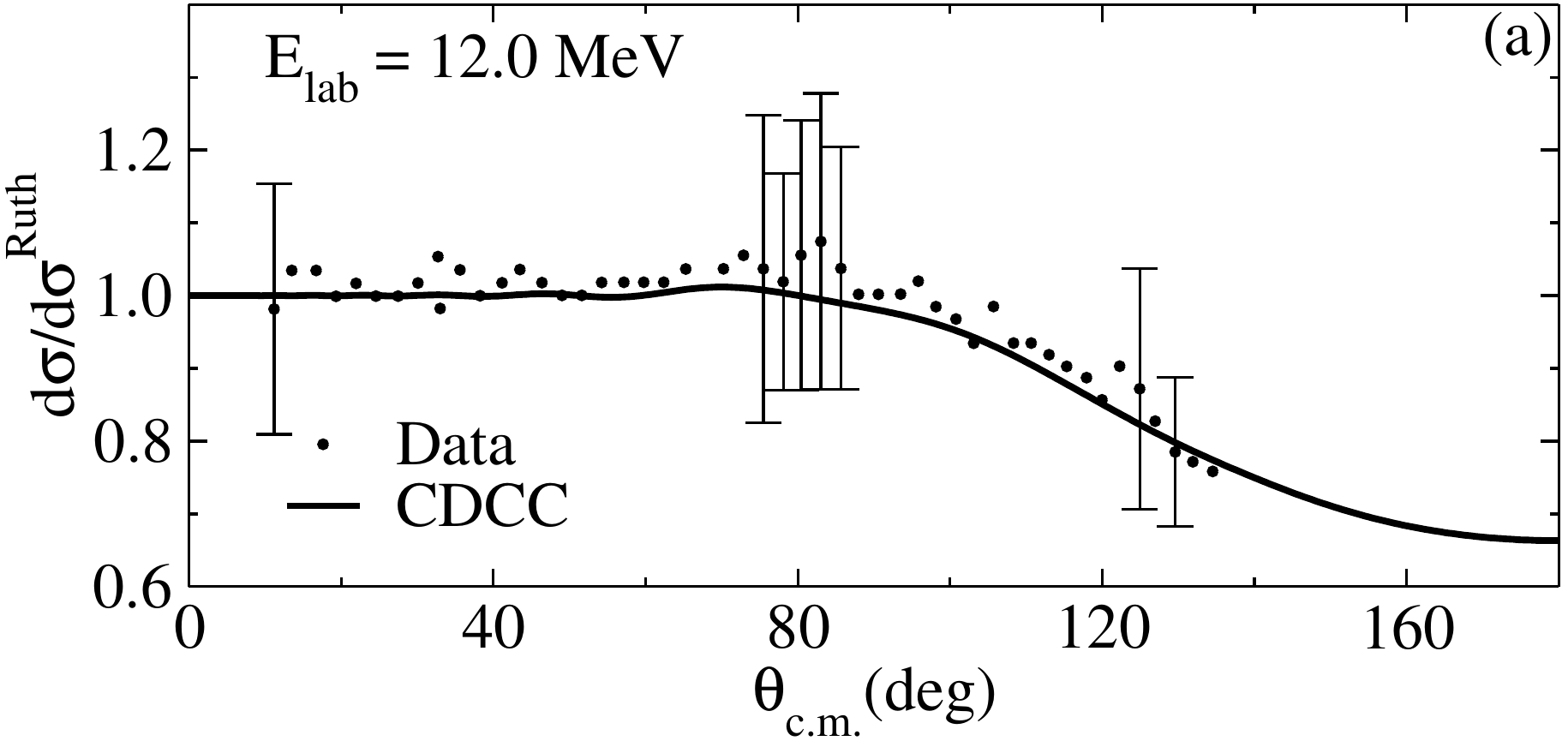}}
 \subfigure{\includegraphics[width=0.8\linewidth]{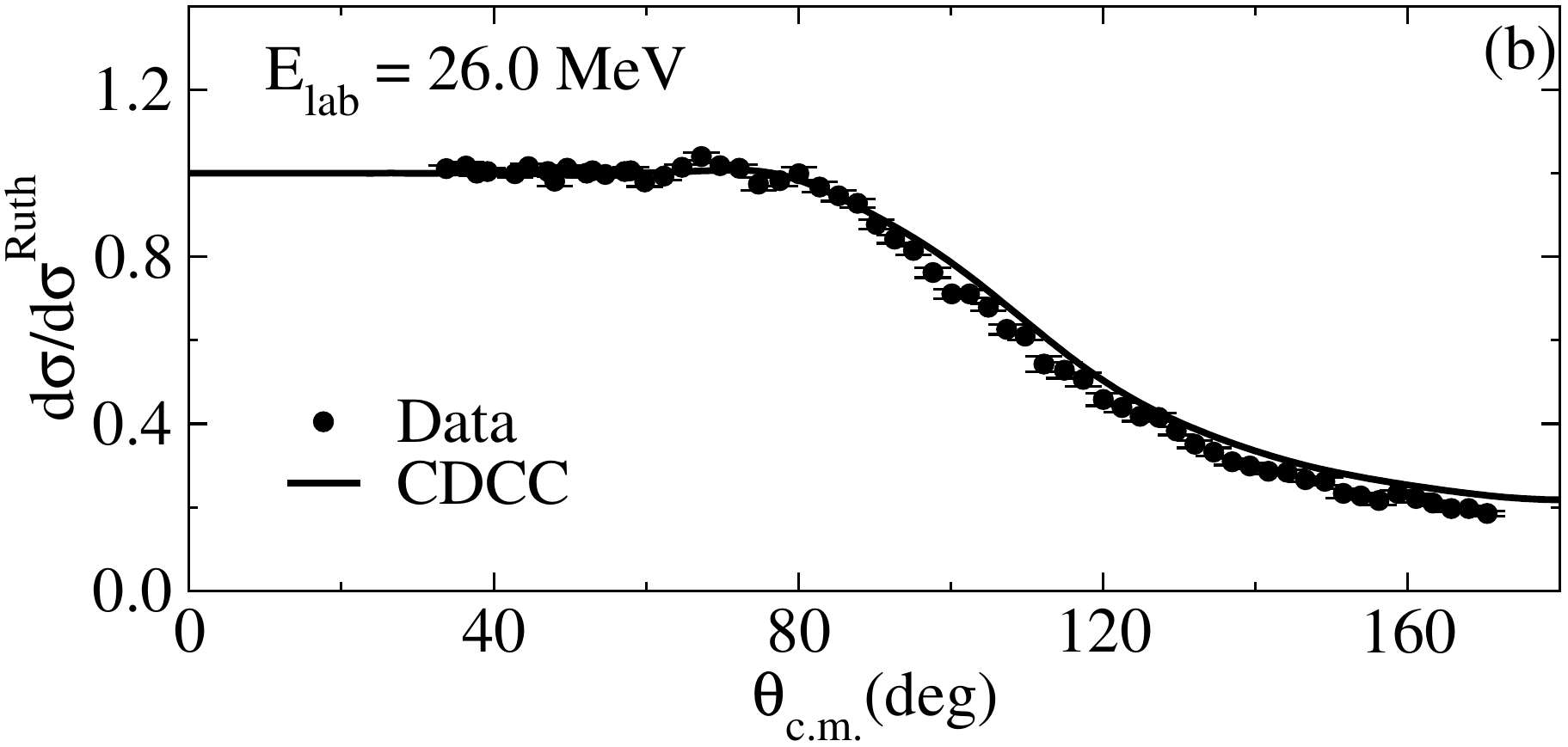}}
 \subfigure{\includegraphics[width=0.8\linewidth]{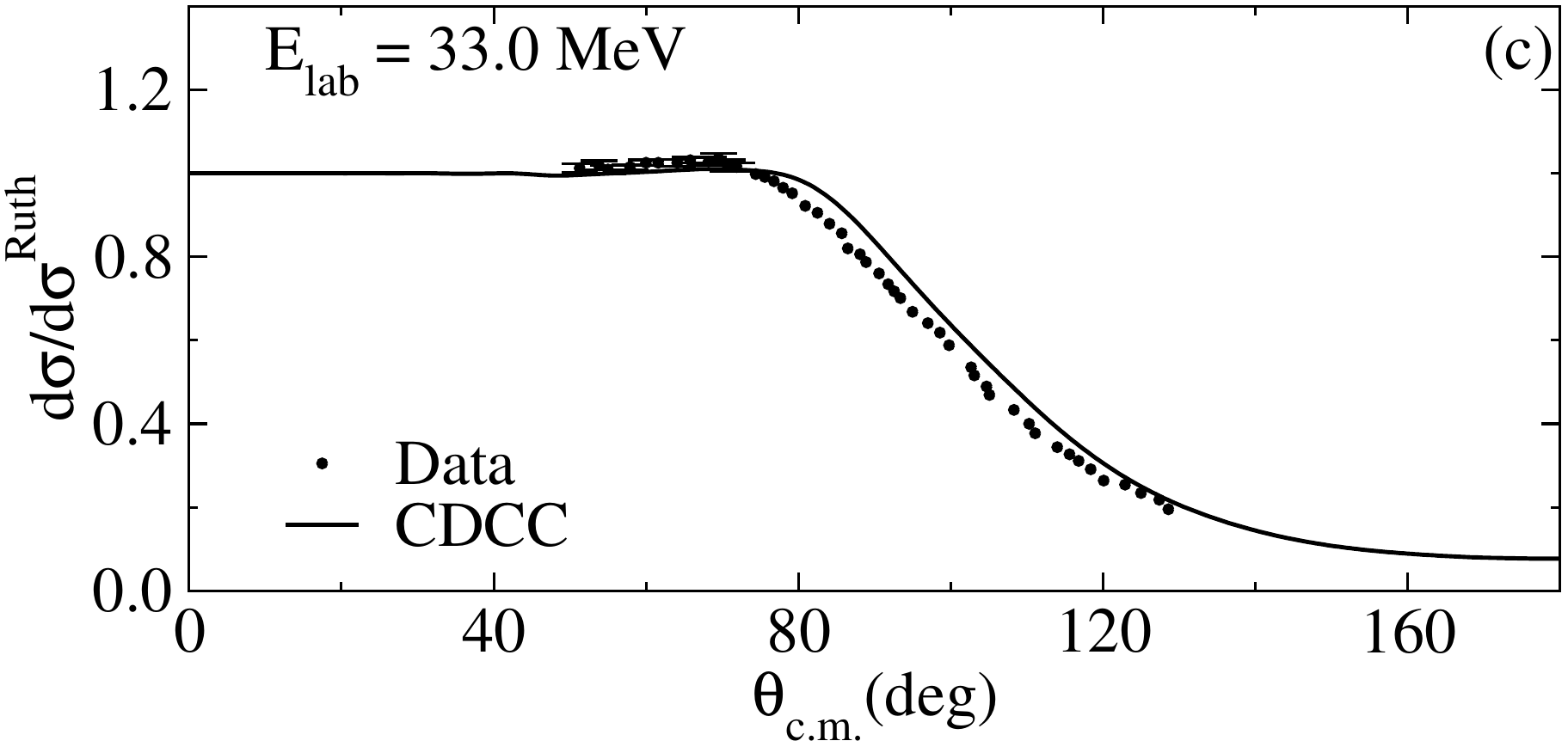}}
 \caption{Elastic scattering angular distributions for the three systems investigated. The data 
 for the $^{59}$Co (panel (a)), $^{144}$Sm (panel (b)) and $^{208} $Pb (panel (c)) targets are 
 respectively from Refs.~\cite{Beck}, 
 \cite{FFA10} and \cite{ANU?}.}
 \label{figure1}
 \end{minipage}
\end{figure}

In this section, we present the results of our CDCC calculations for the breakup of $^6$Li 
projectiles, in collisions with $^{59}$Co, $^{144}$Sm and $^{208}$Pb targets. As a preliminary 
step, we checked that the predictions of the method for the elastic scattering of these systems 
were in good agreement with the available data. The situation is illustrated in figure 
\ref{figure1}, which compares theory and experiment at one near-barrier energy for each system. 
The good agreement obtained in the case of elastic scattering justifies the use of the model in 
calculations of other processes.

\medskip

\begin{figure}[ht]
\centering
 \begin{minipage}[ht]{1\linewidth}
 \centering
 \subfigure{\includegraphics[height=0.7\linewidth]{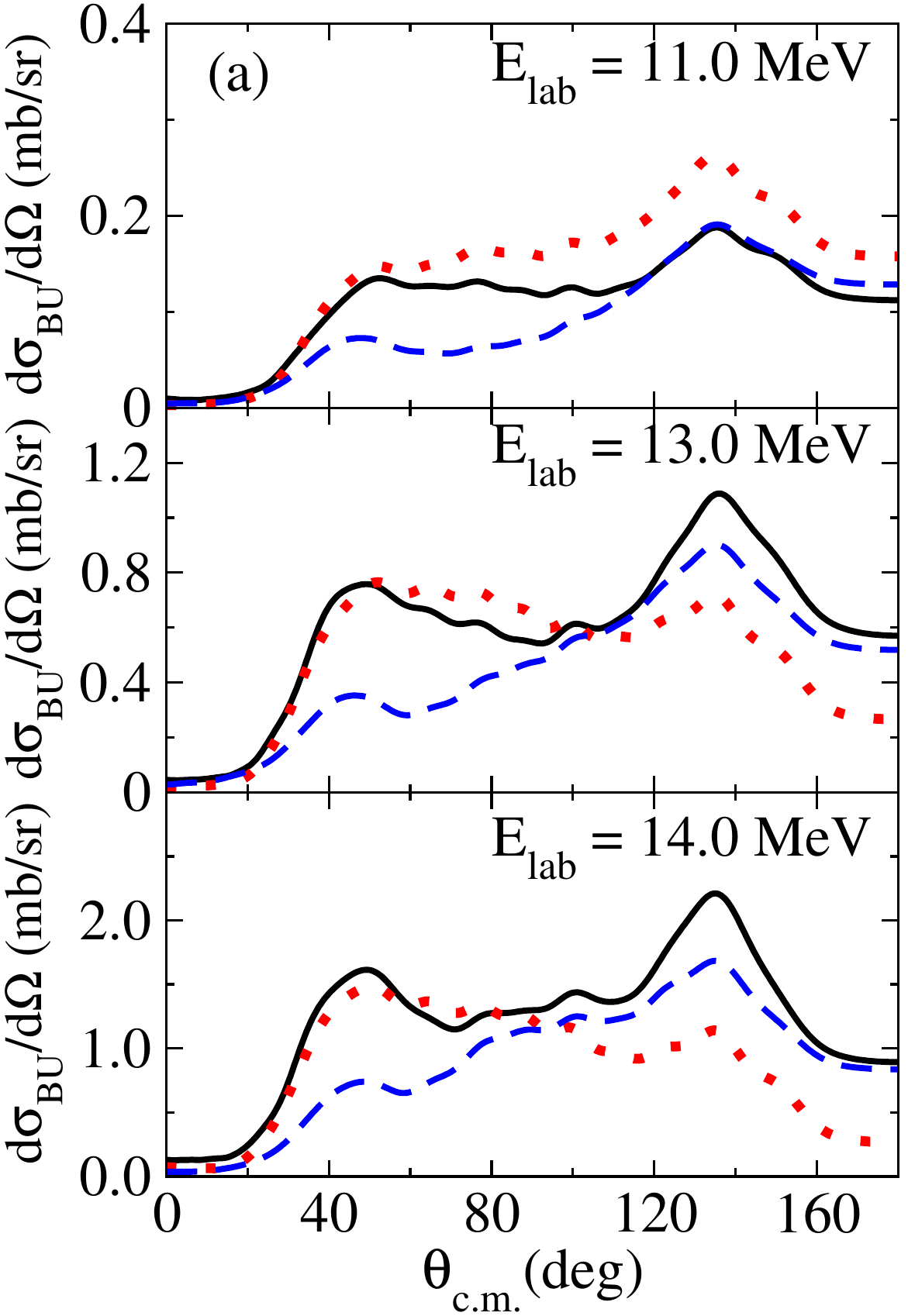}}
 \subfigure{\includegraphics[height=0.7\linewidth]{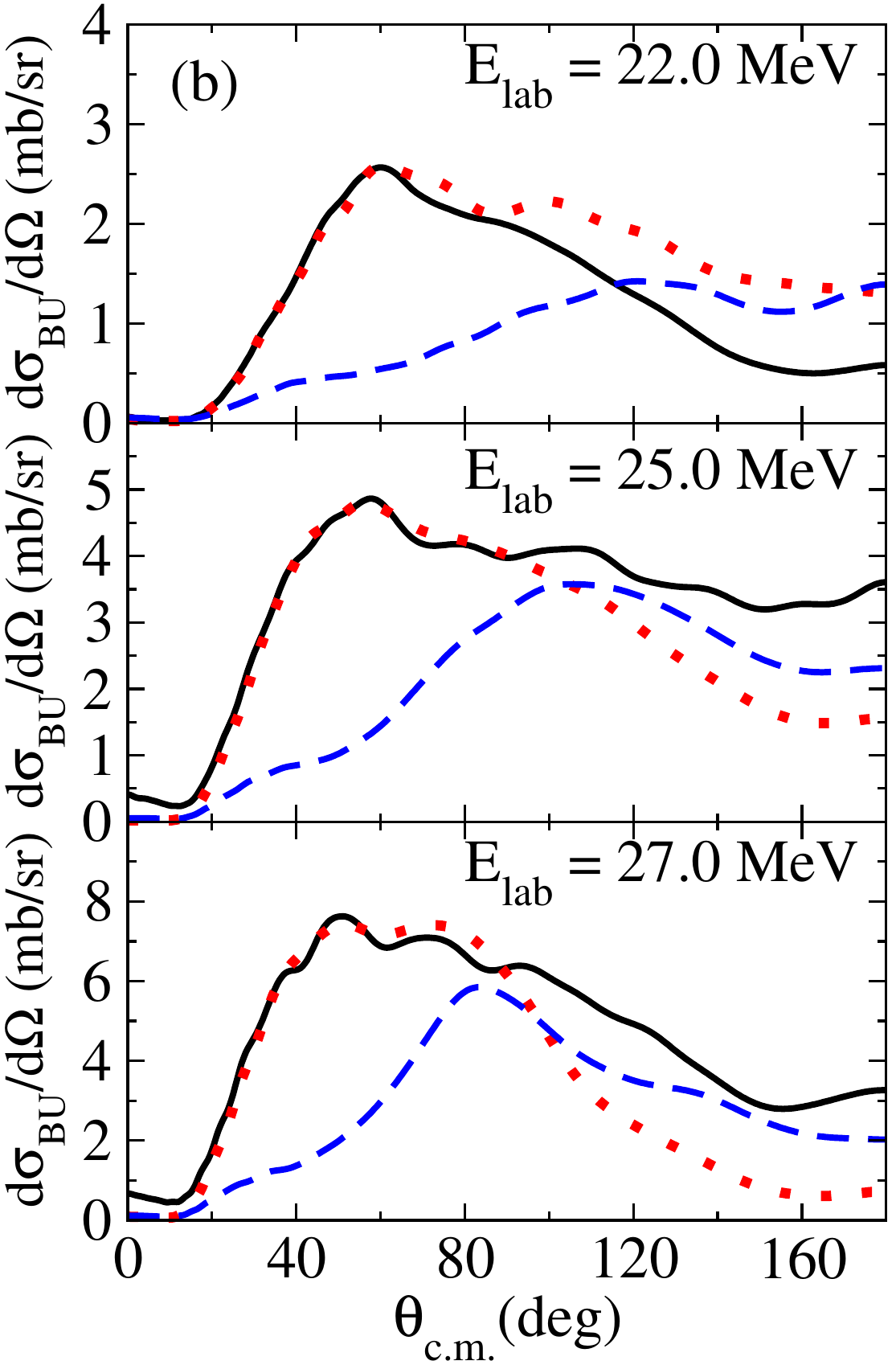}}
 \subfigure{\includegraphics[height=0.7\linewidth]{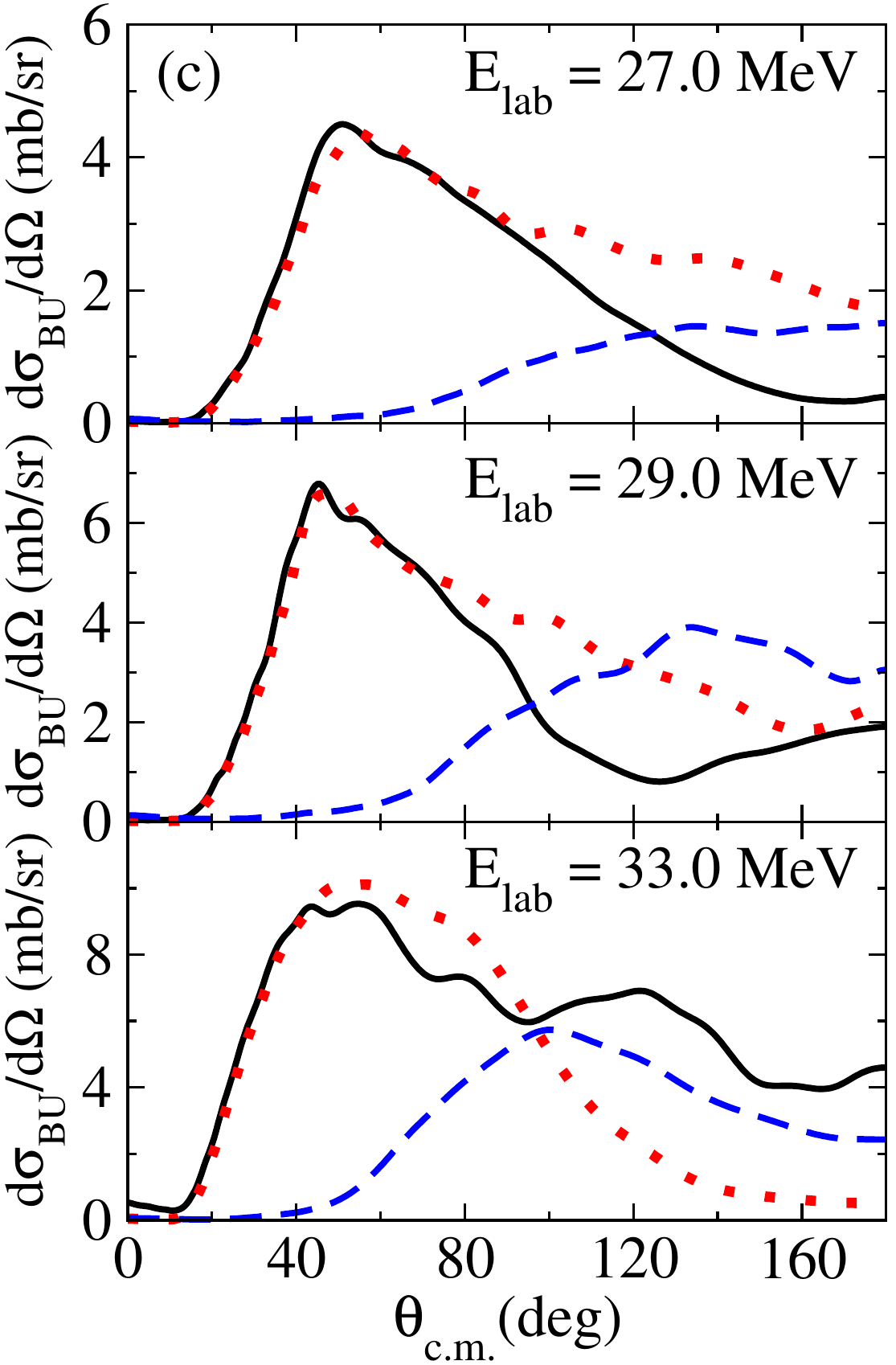}}
 \caption{(Color online) Breakup angular distributions at 3 near-barrier energies for the systems discussed
 in the text. In each case, we show the total breakup cross section (solid lines) together with 
 the cross sections for pure Coulomb (short-dashed lines) and pure nuclear (long-dashed lines) 
 breakup. As in the previous figure, panels (a), (b) and (c) correspond respectively
 to results for the $^{59}$Co, $^{144}$Sm and $^{208} $Pb targets.}
 \label{figure2}
 \end{minipage}
\end{figure}

\subsection{Angular distributions in $^6$Li breakup}

We now turn to the calculations of breakup cross sections, with the purpose of assessing the 
relative importance of the Coulomb and the nuclear contributions to the breakup process. We begin 
with a study of angular distributions. In figure \ref{figure2}, we show the breakup cross 
sections at 3 near-barrier energies for each of the systems mentioned above. The figures exhibit 
the total breakup cross section, together with the separate contributions from Coulomb and nuclear 
breakup. The systems and the collision energies are given inside each subfigure. As a 
reference, we mention that the heights of the Coulomb barrier calculated with the S\~ao Paulo 
potential are $V_{\scr B}\left( ^{59}{\rm Co} \right) = 12.8$ MeV, 
$V_{\scr B}\left( ^{144}{\rm Sm} \right) = 23.8$ MeV and $V_{\scr B}\left( ^{208}{\rm Pb} \right) 
= 29.4$ MeV. Note that for each system the lowest energy in the figure is below the barrier 
whereas the highest is above it. Inspecting figure \ref{figure2}, we note that at the lowest 
energies the Coulomb contribution tends to dominate. In fact, for energies much below the barrier
one expects that only Coulomb breakup can take place, owing to the short-range of the nuclear 
forces. On the other hand, at energies above the barrier Coulomb breakup dominates in the breakup
at forward angles whereas nuclear breakup tends to dominate at large angles. This is not 
surprising, since large angle scattering corresponds to small impact parameters, for which the 
trajectories reach small projectile-target separations, within the reach of the nuclear forces. 

\begin{figure}[h] 
\centering 
\includegraphics[width=8.5cm]{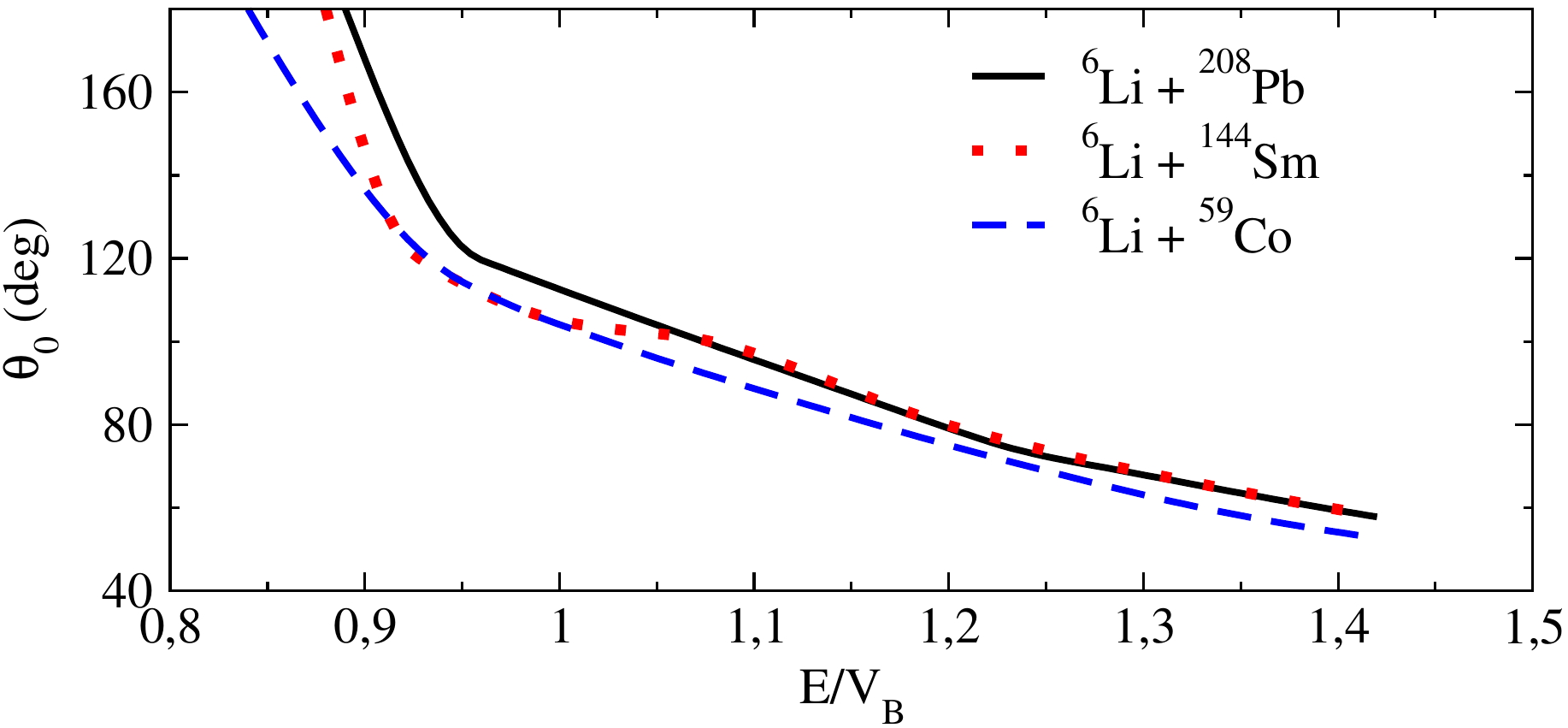} 
\caption{(Color online) The crossing angle, $\theta_{0}$, above which nuclear breakup dominates the breakup
angular distribution. For details see the text.} 
\label{cross} 
\end{figure}

The transition from the Coulomb dominated angular region to the nuclear dominated one occurs at some 
crossing angle, $\theta_0$, which is a function of the collision energy. We have investigated the 
energy dependence of this angle for the three systems of figure \ref{figure2}. The results are 
given in figure \ref{cross}, which shows the crossing angle as a function of the collision energy, 
normalised with respect to the barrier height. First, we notice that the crossing angle grows 
monotonically as the collision energy decreases. In this way, there is a critical energy below
which Coulomb breakup dominates at all angles. Thus, in this energy region there is no crossing 
angle. The second interesting feature of this figure is that the results have a very weak 
dependence on the target. Thus, $\theta_0$ can be approximated as a single function of 
$E / V_{\scr B}$, even for targets in different mass ranges.

\subsection{Integrated breakup cross sections}

We now investigate the total breakup cross sections, integrating the angular distributions of the 
previous sub-section over angles. Results for the systems of the previous figures at three 
collision energies are given in table \ref{tabela}.  The three energies for the different systems 
correspond to approximately the same values of E$_{c.m.}$/V$_{\scr B}$. 

\begin{table}[h]
 \begin{center}
  \begin{tabular}{c|c|c|c|c}
  \hline
  \multicolumn{5}{c}{$^6$Li + $^{59}$Co} \\ 
  \hline
  $\ \ E_{\scr lab}\ \ $  & $\ \ \sigma^{\scr BU}_{\scr Nuc}\ \ $   & 
  $\ \ \sigma^{\scr BU}_{\scr Cou}$ \ \    & \ \ $\sigma^{\scr BU}_{\scr tot}$ \ \  & \ \ ($\sigma^{\scr BU}_{\scr tot}$ - $\sigma^{\scr BU}_{\scr Nuc}$) / $\sigma^{\scr BU}_{\scr Cou}$ \ \    \\ 
  \hline
  11.0 & 0.84 & 1.44 & 1.11 & 0.19 \\ 
  \hline
  13.0 & 4.33 & 5.31 & 5.68 & 0.25 \\ 
  \hline
  14.0 & 8.72 & 9.27 & 11.56 & 0.31 \\ 
  \hline
  \multicolumn{5}{c}{$^6$Li + $^{144}$Sm} \\ 
  \hline
  $\ \ E_{\scr lab}\ \ $  & $\ \ \sigma^{\scr BU}_{\scr Nuc}\ \ $   & 
  $\ \ \sigma^{\scr BU}_{\scr Cou}$ \ \    & \ \ $\sigma^{\scr BU}_{\scr tot}$ \ \  & \ \ ($\sigma^{\scr BU}_{\scr tot}$ - $\sigma^{\scr BU}_{\scr Nuc}$) / $\sigma^{\scr BU}_{\scr Cou}$ \ \    \\ 
  \hline
  22.0 & 11.3 & 22.1 & 18.8 & 0.34 \\ 
  \hline
  25.0 & 30.0 & 41.6 & 48.0 &  0.43 \\ 
  \hline
  27.0 & 43.6 & 57.3 & 69.6 & 0.45 \\ 
  \hline
  \multicolumn{5}{c}{$^6$Li + $^{208}$Pb} \\ 
  \hline
  $\ \ E_{\scr lab}\ \ $  & $\ \ \sigma^{\scr BU}_{\scr Nuc}\ \ $   & 
  $\ \ \sigma^{\scr BU}_{\scr Cou}$ \ \    & \ \ $\sigma^{\scr BU}_{\scr tot}$ \ \  & \ \ ($\sigma^{\scr BU}_{\scr tot}$ - $\sigma^{\scr BU}_{\scr Nuc}$) / $\sigma^{\scr BU}_{\scr Cou}$ \ \    \\ 
  \hline
  27.0 & 8.8 & 34.9 & 29.3 & 0.58 \\ 
  \hline
  29.0 & 22.8 & 46.8 & 37.2 & 0.31 \\ 
  \hline
  33.0 & 38.7 & 66.8 & 82.5 & 0.66 \\ 
  \hline
  \end{tabular}
 \end{center}
 \caption{Integrated breakup cross section for the systems discussed in the text, for three 
 collision energies. The energies are given in MeV and the cross sections in mb.}
 \label{tabela}
\end{table}

Table \ref{tabela} shows some interesting properties of the breakup cross sections. First, one 
notices that, as expected, the Coulomb breakup cross sections are systematically larger for 
heavier systems. On the other hand, the strongest nuclear breakup occurs for the intermediate 
mass $^{144}$Sm target. This is probably associated with the distance of closest approach and 
the width of the effective barrier in this collision but a quantitative interpretation of this behaviour
would require further study. An important remark is that adding the nuclear with the Coulomb breakup 
cross sections, one does not get the total breakup cross section. For example, in the case of 
$^6$Li breakup with the $^{208}$Pb target at 29 MeV, the Coulomb breakup cross section alone is 
larger than the total breakup cross section. The same happens for the other systems at the lowest
collision energy (below the barrier). This is a consequence of destructive Coulomb-nuclear 
interference in the breakup process. Note that this effect can also be observed in the angular 
distributions (e.g. figure \ref{figure2} for the $^{208}$Pb target at 29 MeV, around 
$\theta\sim 120$ degrees). The last column of table \ref{tabela} shows the ratio  ($\sigma^{\scr BU}_{\scr tot}$ - 
$\sigma^{\scr BU}_{\scr Nuc}$) / $\sigma^{\scr BU}_{\scr Cou}$. Note that this ratio is always less than one. 
This fact clearly shows the destructive character of the nuclear-Coulomb interference. 

\begin{figure}[h]
\centering 
\includegraphics[width=8.5cm]{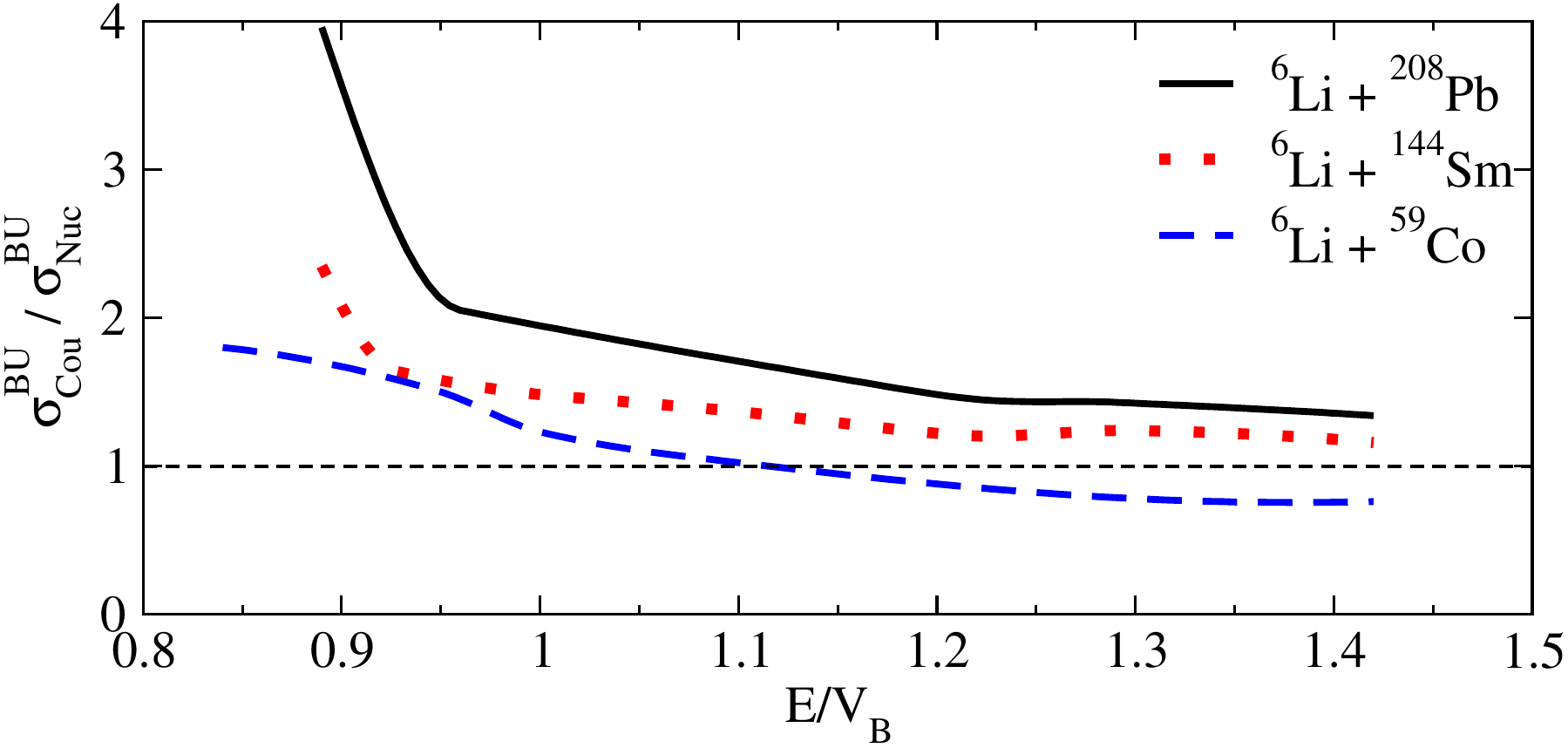} 
\label{fig6} 
\caption{(Color online) Coulomb to nuclear ratio of integrated breakup cross section for the three systems under 
investigation.} 
\label{CNration}
\end{figure}

A more systematic study of the relative importance of the Coulomb and the nuclear forces
in the breakup process is presented in figure \ref{CNration}, which shows the ratio 
$\sigma^{\scr BU}_{\scr Cou} / \sigma^{\scr BU}_{\scr Nuc}$, as a function of $E / V_{\scr B}$.
One can observe that near the Coulomb barrier this ratio is larger than unity for the three
systems. However, the relative importance of the Coulomb breakup decreases as the energy
increases. In the case of the lightest target, the ratio becomes smaller than one for 
$E/V_{\scr B} >1.1$. As expected, for the same values of $E/V_{\scr B}$ this ratio increases with 
the target's charge. However, the difference between the Co and Sm targets, at the same relative 
energies, is very small.

\section{Scaling of the breakup cross sections}

An interesting feature of the nuclear breakup cross section is the so-called ``scaling" law, which 
says that this quantity at high enough energies (several tens of MeV/nucleon) should depend on 
the mass number of the target nucleus as \cite{AHPS80, HLNT06}.

\begin{equation}
\sigma^{\scr BU}_{\scr Nuc} = P_{\scr 1} + P_{\scr 2} \ A_{\scr T}^{1/3},
\end{equation}
where the parameters $P_{\scr 1}$ and $P_{\scr 2}$ depend on the projectile, the
structure of the target and the bombarding energy. Ref.~\cite{HLNT06} 
gives results of detailed CDCC calculations of elastic scattering and
breakup cross sections for halo and non-halo systems. They study collisions
of $^{8}$B (one proton-halo nucleus), $^{11}$Be (one neutron-halo nucleus) 
and $^{7}$Be (normal, non-halo nucleus) projectiles, from several target nuclei.
It was found that the above scaling law works best for the nuclear breakup of the
normal  $^{7}$Be projectile. For the halo projectiles, the scaling law works only approximately.
In fact, the cross sections for these projectiles show a maximum for targets of intermediate 
mass, decreasing for heavy targets such as $^{208}$Pb. \\

In the present work we test the scaling law at lower energies, close to the Coulomb barrier. 
Plotting the nuclear breakup cross section as a function of A$_{\scr T}^{\scr 1/3}$, for the 
same bombarding energy, as it was done for high energies, one finds that they decreases 
with the target mass. This is because the cross section changes rapidly as the collision
energy reaches the Coulomb barrier. To eliminate this effect, we normalise the
collision energy with respect to the Coulomb barrier. That is, we compare cross
sections for the same value of $E_{\rm c.m.}/V_{\scr B}$. 

In figure 5, we show the nuclear breakup cross sections of $^{6}$Li as functions of $A_{\scr T}^{\scr 1/3}$.
The results are for $ E_{\rm c.m.}/V_{\scr B} $= 0.84 (panel (a)), 1.00 (panel (b)), and 1.07 (panel (c)). 
The general behavior resembles the high energy results of Ref.~\cite{HLNT06}, for the 
non-halo weakly bound $^{7}$Be nucleus. In fact the almost straight lines that represent the 
curves for $ E_{\rm c.m.}/V_{\scr B} $= 0.84, 1.00, and 1.07 are fitted with $P_{\scr 1}$ = 
-14.76, -62.60 and -49.89 mb, and $P_{\scr 2} $= 4.11, 16.94 and 15.41 mb, respectively. 
The rather large and negative values of $P_{\scr 1}$ are presumably traced back to barrier 
penetration effects, which limit the use of the 
geometrical picture behind the scaling law. On the other hand, at above-barrier energies the 
values of $P_{\scr 2}$, which supply the slopes of the curves, are practically equal. Accordingly, the 
modified scaling law presented here should supply a useful and easy way to estimate the nuclear 
breakup cross section at other energies close to the barrier, and for other target nuclei.\\

\begin{figure}[t]
\centering 
\includegraphics[width=8.0cm]{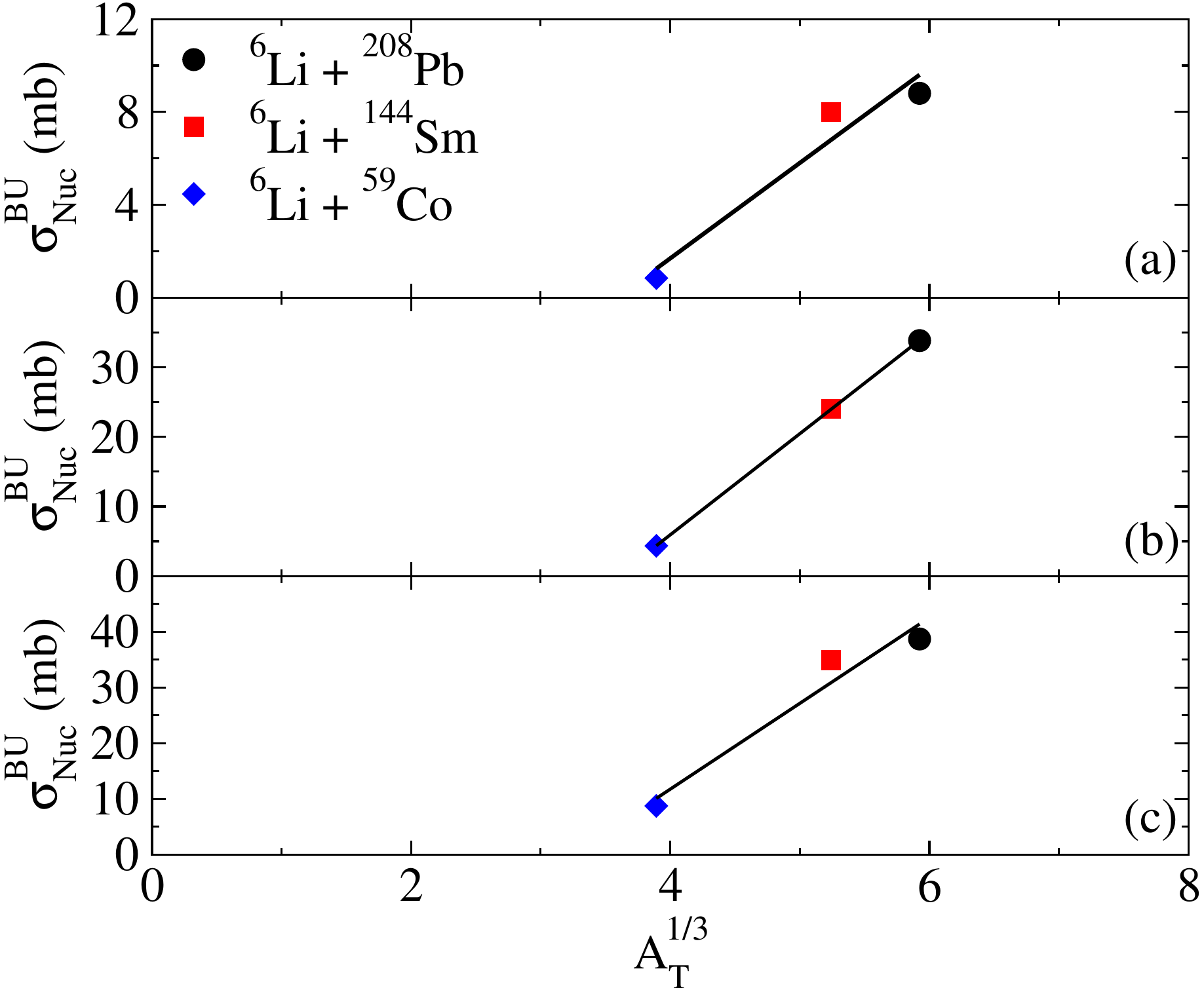} 
\label{fig7}  
\caption{(Color online) Scaling of the nuclear breakup cross sections as a function of $A_{\scr T}^{1/3}$ for 
the systems discussed in the text,  at $E_{\rm c.m.}/V_{\scr B} = 0.84$ (panel (a)), 1.00 (panel (b)) and 1.07 (panel (c)).}
\end{figure}

\begin{figure}[t]
\centering 
\includegraphics[width=8.0cm]{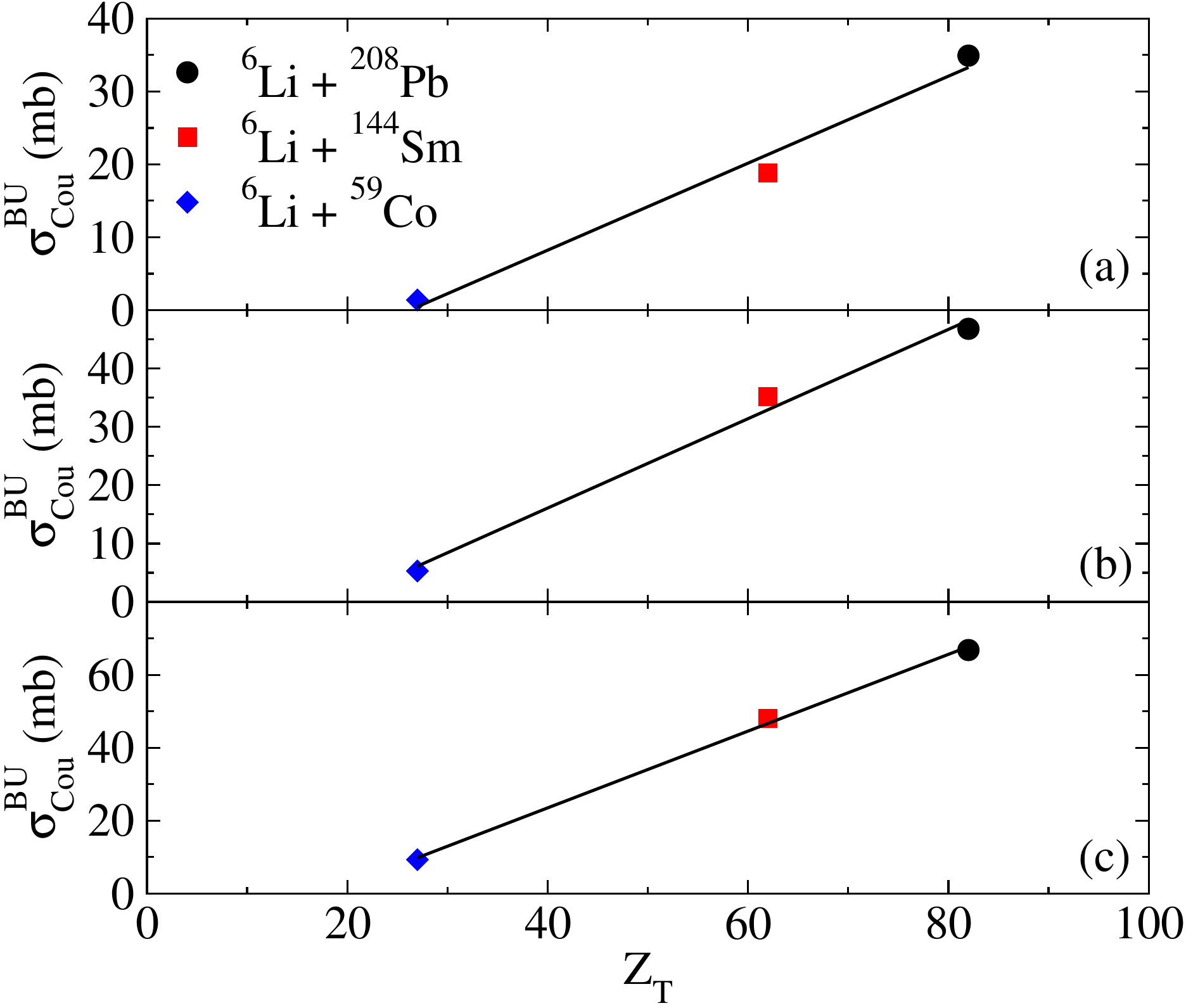} 
\label{Coul-scaling}  
\caption{(Color online) Scaling of the Coulomb breakup cross sections as a function of $Z_{\scr T}$ for the 
systems discussed in the text. Panels (a), (b) and (c) correspond respectively to results at 
$E_{\rm c.m.}/V_{\scr B} = 0.84$ 1.00 and 1.07.}
\end{figure}

One can also derive a scaling law for the Coulomb breakup cross section. In figure 6, we plot 
$\sigma^{\scr BU}_{\scr Cou}$ vs. $Z_{\scr T}$,  for the three systems discussed in the text.
Panels (a), (b) and (c) correspond respectively to results for $E/V_{\scr B} = 0.84, 1.00$ and 1.07.
The figures show that the Coulomb breakup cross sections depend linearly on $Z_{\scr T}$, to
a very good approximation. This behaviour can be qualitatively explained as follows. First, we point
out that the electromagnetic coupling matrix-elements are proportional to $Z_{\scr T}$, which leads
to a $Z_{\scr T}^2$ dependence in the Coulomb breakup cross section. On the other hand, the
cross sections for reaction channels are proportional to a $1/E_{\rm c.m.}$ factor~\cite{wongnew}.
Since in each panel the collision energy corresponds to the same $E_{\rm c.m.}/V_{\scr B}$ ratio,
and $V_{\scr B}$ is roughly proportional to $Z_{\scr T}$, one gets a $1/Z_{\scr T}$ factor. The
combination of the two arguments presented above leads to the linear dependence obtained
in figure 6.

\section{Comparison between fusion and breakup cross sections}

\begin{figure}[t]
\centering \includegraphics[width=8 cm]{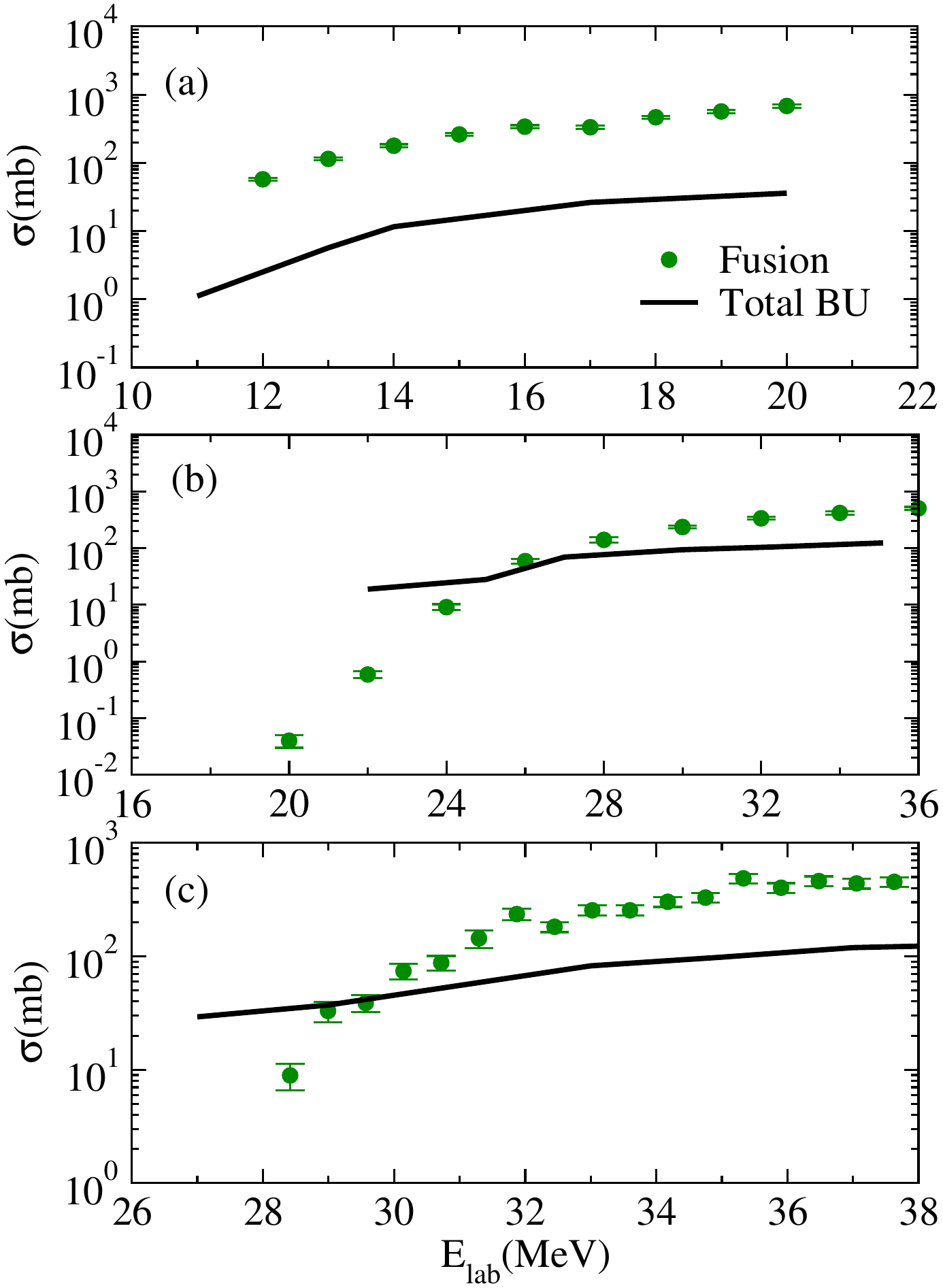} 
\label{fusion-break}  
\caption{(Color online) Comparisons of the fusion and the breakup cross sections for collisions of $^6$Li
projectiles with the $^{59}$Co (panel (a)), $^{144}$Sm (panel (b)) and $^{208} $Pb (panel (c)) targets,
at near-barrier energies.}
\end{figure}

Now we compare the predicted  integrated breakup cross sections with the experimental values of 
the corresponding fusion cross section. We consider the same systems of the previous section, and 
use the fusion data of  Refs.~\cite{beck-fus,rat,wu}. 
Figure 7 
shows these comparisons. First, one notices that the breakup cross section for the light 
$^6$Li + $^{59}$Co system is nearly two orders of magnitude smaller than the fusion cross section. 
Another interesting point is that the ratio of these cross sections is roughly constant in the 
whole energy interval of the figure. For the medium-mass and heavy systems, $^6$Li + $^{144}$Sm 
and $^6$Li + $^{208}$Pb, the situation is different. The breakup cross section is dominant at low
energies whereas the fusion cross section dominates at high energies. The transition takes place 
in the neighbourhood of the Coulomb barrier. This behaviour is due to the Coulomb contribution to
breakup. As the energy decreases below the barrier, the fusion process falls off exponentially. 
This is because fusion takes place when the projectile tunnels through the potential barrier. On 
the other hand, the decrease of Coulomb breakup is much slower. As the collision energy decreases, 
the classical turning point increases. Thus, the Coulomb part of the breakup coupling becomes 
weaker. However, the electromagnetic interactions  at long distances fall as $r^{-(\lambda +1)}$ (where $\lambda$ is the multipolarity of the interaction), which is much slower than the exponential fall off.

\section{Conclusions}

We have investigated the nature of $^6$Li breakup in near-barrier energy collisions, with targets 
in different mass ranges ($^{50}$Co, $^{144}$Sm and $^{208}$Pb).  Our theoretical cross sections 
are based on CDCC calculations, which were shown to lead to accurate predictions of the 
available scattering data. 

For each system, we studied the importance of contributions from the 
nuclear and from the Coulomb fields. We found that at low enough energies 
($E_{\rm c.m.}<0.9\,V_{\scr B}$), the breakup process is mainly due to Coulomb forces, for any
scattering angle. In this energy region, the nuclear and the Coulomb amplitudes for the breakup 
process interfere destructively. In this way, the cross section for pure Coulomb breakup may be larger
than the cross section arising from the simultaneous action of the Coulomb and the nuclear fields. 
Above the Coulomb barrier, Coulomb breakup was shown to dominate at forward 
angles whereas nuclear breakup is dominant at larger angles. The transition takes place at some 
angle $\theta_0$, which increases with decreasing energies. A study of the energy dependence of 
the transition angle indicated that it becomes nearly system independent if plotted as a function 
of the energy normalised with respect to the Coulomb barrier. 

We have shown that the nuclear breakup cross section has a nearly linear dependence of 
$A_{\scr T}^{\scr 1/3}$ as suggested in Ref.~\cite{HLNT06}, for collisions at higher energies. On the 
other hand, the Coulomb breakup cross section was shown to depend linearly of the target charge, 
as could be predicted by qualitative arguments.

Finally, we made a comparison of our theoretical breakup cross sections with the available fusion 
data. We concluded that the latter are about two orders of magnitude larger than the former for 
the light $^6$Li + $^{50}$Co system. For heavier systems, the breakup cross section becomes more
important below the Coulomb barrier.\\

\noindent \textbf{Acknowledgements}
 
\medskip \noindent 
The authors would like to thank the financial support from CNPq, CAPES, FAPERJ, FAPESP and the 
PRONEX.


\begin{thebibliography}{99} 

\bibitem{CGD06} L.F. Canto, P.R.S. Gomes, R. Donangelo, M.S. Hussein, Phys. Rep. \textbf{424}, 
1 (2006). 

\bibitem{CGL09} L.F. Canto, P.R.S. Gomes, J. Lubian, L.C. Chamon, E. Crema; J. Phys. G 
\textbf{36}, 015109 (2009); Nucl. Phys. A \textbf{821}, 51 (2009). 

\bibitem{Gomes09} P. R. S. Gomes, J. Lubian, L. F. Canto, Phys. Rev. C \textbf{79}, 027606 (2009). 

\bibitem{Hussein} M.S. Hussein, P. R. S. Gomes, J. Lubian, L. C. Chamon, Phys. Rev. C \textbf{73}, 
044610 (2006). 

\bibitem{Luong} D. H. Luong \textit{et al.}, Phys. Lett. B \textbf{695}, 105 (2011). 

\bibitem{Dasgupta10} M. Dasgupta \textit{et al.}, Nucl. Phys. A \textbf{834}, 147c (2010). 

\bibitem{Rafiei} R. Rafiei \textit{et al.}, Phys. Rev. C \textbf{81}, 024601 (2010). 

\bibitem{Shiravasta} A. Shrivastava \textit{et al.}, Phys. Lett. B \textbf{633}, 463 (2006). 

\bibitem{KYI86} M. Kamimura, M. Yahiro, Y. Iseri, Y. Sakuragi, H. Kameyama, and M. Kawai, Prog. 
Theor. Phys. Suppl. \textbf{89}, 1 (1986). 

\bibitem{AIK87} N. Austern, Y. Iseri, M. Kamimura, M. Kawai, G. Rawitscher, and M. Yahiro, Phys. 
Rep. \textbf{154}, 125 (1987). 

\bibitem{Otomar10} D. R. Otomar, J. Lubian, P.R.S. Gomes, Europ. Phys. J. A \textbf{46}, 285 (2010). 

\bibitem{SYK82} Y. Sakuragi, M. Yahiro, and M. Kamimura, Prog. Theor. Phys. \textbf{68}, 322 (1982). 

\bibitem{KRu96} N. Keeley and K. Rusek, Phys. Lett. B \textbf{ 357}, 9 (1996). 

\bibitem{Tho88} I. J. Thompson, Comput. Phys. Rep. \textbf{7}, 3 (1988). 

\bibitem{DTB03} A. Diaz-Torres, I. J. Thompson and C. Beck, Phys. Rev. C \textbf{68}, 044607 (2003). 

\bibitem{OLG09} D. R. Otomar \textit{et al.}, Phys. Rev. C \textbf{80}, 034614 (2009). 

\bibitem{Chamon} L.C. Chamon \textit{et al.}, Phys. Rev. Lett. \textbf{79}, 5218 (1997); Phys. 
Rev. C \textbf{66}, 014610 (2002). 


\bibitem{Ram01} S. Raman, C. W. Nestor, Jr., and P. Tikkanena, At. Data, Nucl. Data Tables 
\textbf{78}, 1 (2001). 

\bibitem{Kib02} T. Kib\'{e}di and R. H. Spear, At. Data, Nucl. Data Tables \textbf{80}, 35 (2002). 

\bibitem{pb208} M.J Martin, Nucl. Data Sheets \textbf{108}, 1583 (2007). 

\bibitem{co59} C.M. Baglin, Nucl. Data Sheets \textbf{95}, 215 (2002).

\bibitem{Beck} F.A Souza \textit{et al.}, Phys. Rev. C \textbf{75}, 044601 (2007).

\bibitem{FFA10} J.M. Figueira \textit{et al.}, Phy. Rev. C \textbf{81}, 024613 (2010). 

\bibitem{ANU?} N. Keeley \textit{et al.} Nucl. Phys. A \textbf{571},  326 (1994).

\bibitem{AHPS80} J. C. Acquadro, M. S. Hussein, D. Pereira, and O. Sala, Phys. Lett. B 
\textbf{100}, 381(1980). 

\bibitem{HLNT06} M. S. Hussein, R. Lichtenthaler, F. M. Nunes, and I. J. Thompson, Phys. Lett. B 
\textbf{640}, 91 (2006).

\bibitem{wongnew} C. Y. Wong,  Phy. Rev. C \textbf{86}, 064603 (2012).

\bibitem{beck-fus} C. Beck \textit{et al.}, Phys. Rev. C, \textbf{67}, 054602 (2003).

\bibitem{rat} P.K. Rath \textit{et al.}, Phys. Rev. C, \textbf{79}, 051601 (2009).

\bibitem{wu}  Y.W. Wu \textit{et al.}, Phys. Rev. C, \textbf{68} 044605 (2003).

 \end{thebibliography}
\end{document}